\documentclass[preprint,aps,amsmath,superscriptaddress,nofootinbib]{revtex4}
\usepackage{bm}
\usepackage{epsfig}
\usepackage{epstopdf}
\usepackage{graphics,graphicx}
\usepackage{fancyhdr,fancybox}
\usepackage{feynmf}
\usepackage[small]{caption}

\newcommand{\nn}{\nonumber} 

\newcommand{\bea}{\begin{eqnarray}}
\newcommand{\eea}{\end{eqnarray}}

\newcommand{\Hb}{\bar{H}}
\newcommand{\Pb}{\bar{P}}
\newcommand{\Vb}{\bar{V}}

\begin{document}



\title{Hadronic Loops versus Factorization in Effective Field Theory  calculations of $X(3872) \to \chi_{cJ} \pi^0$ }

\author{Thomas Mehen\footnote{Electronic address: mehen@phy.duke.edu}}
\affiliation{Department of Physics, 
	Duke University, Durham,  
	NC 27708\vspace{0.2cm}}


\date{\today\\ \vspace{1cm} }


\begin{abstract}
We compare two existing approaches to calculating the decay of molecular quarkonium states to conventional quarkonia in effective field theory,
using $X(3872) \to \chi_{cJ} \pi^0$ as an example. In one approach the decay of the molecular quarkonium proceeds through
a triangle diagram with charmed mesons in the loop. We argue this approach predicts excessively large rates for $\Gamma[X(3872) \to \chi_{cJ}\pi^0]$ 
unless both charged and neutral mesons are included and a cancellation between these contributions is arranged to suppress the decay rates. This cancellation 
occurs naturally if the $X(3872)$ is primarily in the $I=0$ $D \bar{D}^{*} +c.c.$ scattering channel. The factorization approach to molecular decays calculates the rates in terms of tree-level transitions for the $D$ mesons in the $X(3872)$ to the final state, multiplied by unknown matrix elements. We show that this approach is equivalent to hadronic loops approach if the cutoff on the loop integrations is taken to be a few hundred MeV or smaller, as is appropriate when the charged $D$ mesons have been integrated out of the effective theory.

\end{abstract}

\maketitle

\newpage

\section{Introduction} 

The last ten years have seen a plethora of discoveries of unconventional quarkonia,~\footnote{For a review of recent developments in quarkonium spectroscopy, we refer the reader to Refs.~\cite{Olsen:2014qna,Bian:2014vfa,Liu:2013waa,Brambilla:2010cs} } the first and most studied of these being the $X(3872)$~\cite{Choi:2003ue,Acosta:2003zx, Abazov:2004kp,Aubert:2004ns}. 
Because of its proximity to the $D^0\bar{D}^{*0}$ threshold  it is thought by many
authors to be a molecular state. If the state consists primarily of the $C$ even linear combination of neutral $D$ mesons, 
$D^0 \bar{D}^{*0}$ + c.c., the binding energy is $-0.11 \pm 0.21$ MeV, and this state 
is a very shallow bound state. For the central value of this binding energy, one calculates the 
typical separation of the $D^0$ and $\bar{D}^{*0}$ to be approximately 10 fm, which 
is an astonishingly large length scale compared to typical hadronic scales. Ref.~\cite{Fleming:2007rp} exploited this separation of scales 
to construct an effective field theory for the $X(3872)$  called XEFT.  Heavy hadron chiral perturbation theory (HH$\chi$PT)~\cite{Wise:1992hn,Burdman:1992gh,Yan:1992gz}
is matched onto a non-relativistic theory of neutral $D$ mesons and pions. Their interactions are constrained 
by the heavy quark and chiral symmetries of QCD. A contact interaction is tuned to produce a shallow bound state in the $D^0\bar{D}^{*0} + c.c.$
channel which is the $X(3872)$. The structure of the theory is similar to effective field theories of the deuteron and low energy two-body nuclear physics~\cite{Kaplan:1998tg,Kaplan:1998we}.

For processes that are dominated by long-distance aspects 
of the $X(3872)$, such as $X(3872) \to D^0 \bar{D}^0 \pi^0$ or $D^0 \bar{D}^0 \gamma$,
this theory reproduces effective range theory (ERT) at lowest order. ERT predictions
for these $X(3872)$ decays were first calculated  in Refs.~\cite{Voloshin:2003nt,Voloshin:2005rt}.
XEFT allows for the systematic inclusion of corrections to these predictions
from pion loops and higher dimension operators. Ref.~\cite{Fleming:2007rp} showed the corrections 
from pion loops were negligible, at least for the process $X(3872) \to D^0\bar{D}^0 \pi^0$. The effect of final state interactions on the reaction
$X(3872) \to D^0 \bar{D}^0 \pi^0$ was recently studied in Ref.~\cite{Guo:2014hqa}.
For calculations of many processes within XEFT, see Refs.~\cite{Fleming:2008yn,Canham:2009zq,Braaten:2010mg,Mehen:2011ds,Baru:2011rs, Fleming:2011xa,Margaryan:2013tta}.
XEFT has also been used to calculate the  quark mass dependence of the $X(3872)$ binding energy  in Ref.~\cite{Jansen:2013cba}, for a related EFT calculation see Ref.~\cite{Baru:2013rta}.

Many observations of $X(3872)$ involve decays 
to conventional charmonia, including $X(3872)\to J/\psi \pi^+\pi^-, J/\psi \pi^+\pi^- \pi^0, J/\psi \gamma$, and $\psi(2S) \gamma$.
The $X(3872)$ has also recently been observed  in the decay of the exotic  quarkonium state $Y(4260)  \to X(3872) \gamma$~\cite{Ablikim:2013dyn}. Ref.~\cite{Guo:2013zbw} predicted an enhanced rate for the decay 
$Y(4260) \to X(3872) \gamma$ based on the assumption that the $Y(4260)$ is a $D\bar{D}_1$ molecule, while other
authors interpret the $Y(4260)$ as a charmonium hybrid~\cite{Close:2005iz},  so this transition  
probably does not involve a compact $c\bar{c}$ state. Other
 decay and production processes with conventional charmonia such as $J/\psi$ or $\psi(2S)$ involve short-distance scales since the $D$ and $\bar{D}^*$ 
must coalesce to couple to a conventional  charmonium.  For these decays  there exist  two distinct approaches to applying XEFT in the literature. 
The approach first taken in Ref.~\cite{Fleming:2008yn} is to use  HH$\chi$PT
to calculate the transition of $D^0 \bar{D}^{*0}$ to the final state, then match the resulting amplitudes onto 
XEFT operators. The resulting prediction for the partial decay width of the $X(3872)$ is given  by an expression 
of the form
\bea\label{factorization}
\Gamma[X(3872) \to F.S.] \propto {\cal O}_{XEFT} \times \sigma[D^0\bar{D}^{*0}+c.c. \to F.S.] \, ,
\eea
where $F.S.$  denotes the final state (which includes a charmonium) and ${\cal O}_{XEFT}$ is an XEFT operator. This operator
plays the same role as  the wave function at the origin squared in a traditional approach to bound state
calculations. The numerical value of the XEFT operator  is unknown and must be 
extracted from data. Since the $D^0$ and $\bar{D}^{*0}$ must coalesce to form the compact 
charmonium, part of the process involves short-distance physics that is not determined by the universal 
nature of the long-distance part of the $X(3872)$ wave function, and this physics is encoded in   ${\cal O}_{XEFT}$.
Similar factorization theorems for $X(3872)$ decay and production were developed in Refs.~\cite{Braaten:2005jj,Braaten:2006sy}.
 We will refer to the approach to $X(3872)$ decays 
advocated in Refs.~\cite{Fleming:2008yn,Braaten:2005jj,Braaten:2006sy} which yields a factorized formulae of the form of Eq.~(\ref{factorization}) as the {\em factorization}
approach to $X(3872)$ decays.  

The second EFT approach to $X(3872)$ production and decays is advocated in, e.g., Refs.~\cite{Guo:2013zbw,Guo:2014taa}.
The decay involving the conventional quarkonium proceeds
through a loop diagram in which both the $X(3872)$ and the conventional quarkonium couple to heavy mesons. In this case the $X(3872)$ coupling to heavy mesons in the loop is  fixed by the residue of the pole 
in the $T$-matrix. In some cases \cite{Guo:2013zbw} a power counting argument shows that the hadronic loop is lower order than any tree-level diagram and the hadronic loop approach is more predictive than factorization since there is no undetermined XEFT matrix element. Whether or not this happens depends on the quantum numbers of the states
involved in the transition. For example, in the radiative transitions $J/\psi, \psi(2S)\to X(3872) \gamma$ a counterterm appears 
at leading order, so it is not possible to predict the ratio $\Gamma[\psi(2S)\to X(3872) \gamma]/\Gamma[J/\psi \to X(3872)\gamma]$ \cite{Guo:2014taa}.  A similar conclusion was reached in the factorization approach in Ref.~\cite{Mehen:2011ds}.
Even when there is no tree-level counterterm at leading order,  this approach still may not be entirely predictive since the couplings of heavy mesons to conventional quarkonia in the loop
could be unknown. We will refer to the approach to $X(3872)$ decays in which the decay is assumed to go through a hadronic loop 
as the {\it hadronic loop} approach to calculating $X(3872)$ decays.

In addition to two different approaches to calculating $X(3872)$ decays, there are also different   choices of   the relevant degrees of freedom appropriate for an effective theory suitable for describing the $X(3872)$. In the literature  there are calculations within both the factorization approach and the hadronic loop approach that  only include 
neutral $D$ mesons as explicit degrees of freedom, since the $X(3872)$ is considered a shallow bound state of these mesons alone.~\footnote{For some processes there are other justifications for neglecting the charged mesons. For example, in the calculation of $X(3872)$ radiative decays in Ref.~\cite{Guo:2013zbw}, the charged mesons were neglected because the neutral charmed mesons couple much more strongly to the photon. However, charged mesons are included in  analysis of $X(3872)$ radiative decays in Ref.~\cite{Guo:2014taa}.}
Refs.~\cite{Aceti:2012cb,Aceti:2012qd} have  emphasized the importance of including charged $D$ mesons as well in the calculations of the decays $X(3872) \to J/\psi \gamma,J/\psi \pi^+ \pi^- $, and $J/\psi \pi^+ \pi^- \pi^0$.  Note that the charged meson threshold $D^+D^{*-} + c.c.$ is considerably 
farther away from the $X(3872)$ mass than the neutral threshold. The binding energy is 8.2 MeV and the corresponding estimate of 
the separation of the charged mesons in the $X(3872)$ is 1.1 fm. This is roughly a factor of 10 smaller  than the central value for the  corresponding estimate for the neutral channel. The charged $D$ mesons are separated by a distance that is not much larger than the size of the hadrons themselves. 
At length scales larger than 1 fm, the wavefunction is certainly dominated by  the neutral mesons. In the original 
formulation of XEFT the charged mesons are integrated out of the theory and their effects subsumed in to short-distance XEFT operators. 
However, for processes like decays to conventional charmonium, in which both long and short distance scales are important, it may be 
desirable to include these as explicit degrees of freedom.

The purpose of this paper is to compare the  different approaches to calculating the decay of $X(3872)$ to conventional quarkonia, using the decays $X(3872) \to \chi_{cJ} \pi^0$ as an example.   These decays were first studied in Ref.~\cite{Dubynskiy:2007tj} where it was pointed out that the relative 
rates for different $J$ are predicted by heavy quark symmetry and can be used to distinguish between different interpretations 
of the $X(3872)$. These decays were  studied in XEFT in Ref.~\cite{Fleming:2008yn}, which used factorization in a theory with only neutral $D$ mesons as explicit degrees of freedom.           Ref.~\cite{Fleming:2008yn} showed that within this approach there are two distinct long-distance and short-distance mechanisms contributing to the decay and the relative rates depend on the relative importance of the two mechanisms. The authors of Ref.~\cite{Fleming:2008yn} also computed the partial widths using the hadronic loop formalism, with only neutral $D$ mesons as explicit degrees of freedom,  but as we will see in the next section, this yields exceedingly large partial widths for $X(3872) \to \chi_{cJ} \pi^0$ that are in conflict with experiment, so this approach was discarded and the calculation was not published in Ref.~\cite{Fleming:2008yn}.  This result is somewhat model dependent as the predicted rates depend on the unknown coupling  of the $\chi_{cJ}$ to charmed mesons, which is estimated using the model in  Ref.~\cite{Colangelo:2003sa}. However, to make the predicted partial widths for $X(3872) \to \chi_{cJ} \pi^0$ consistent with experiment requires that this coupling be almost two orders of magnitude smaller than what one expects from naive dimensional analysis. We conclude that the hadronic loop approach with only neutral charmed mesons as explicit degrees of freedom is inconsistent with experiment.
The hadronic loop approach can be made consistent with data if charged mesons are included as explicit degrees of freedom. If the $X(3872)$ has nearly equal couplings  to the charged and neutral channels a cancellation  between charged and neutral loop diagrams  suppresses the rate. This cancellation occurs naturally if the $X(3872)$ is an $I=0$ state. An $I=0$ interpretation of the $X(3872)$ has been put forth by other authors~\cite{Aceti:2012cb,Aceti:2012qd} and is consistent with the observed ratio 
$\Gamma[X(3872) \to J/\psi \pi^+ \pi^-\pi^0]/\Gamma[X(3872) \to J/\psi \pi^+ \pi^- ] = 0.8\pm 0.3$~\cite{Agashe:2014kda}
 if one accounts for differences in two- and three-body phase space~\cite{Colangelo:2007ph}.
  
 In section III, we discuss how the hadronic loops approach is related to the factorization approach. We show that the hadronic loop integral can be expressed as the convolution of the ERT wave function of the $X(3872)$ with the tree-level matrix element for $D^{*0}\bar{D}^0 \to \chi_{cJ}\pi^0$~\footnote{Here and throughout this paper charge conjugate channels are implied.}.
 We simplify the calculation by dropping some terms $O(p_\pi^2/(m_D E_\pi))$ which only changes answers by a few percent. Then the hadronic loop integrals contain contributions from two very different scales, $ \gamma_n = 14$ MeV and $ \sqrt{m_D E_\pi} \approx 850$ MeV. Here  $\gamma_n$ is the binding momentum in the neutral channel, $m_D$ is the $D$ meson mass, and $E_\pi$ is the energy of the pion in the decay. 
 The  contribution from loop momentum of order $\sqrt{ m_D E_\pi }$ gives the dominant contribution to the integral, but we argue that in a theory in which the charged mesons have been integrated out, the theory must be thought of as having a cutoff $\Lambda \sim \gamma_c$, where $\gamma_c$ is the binding momentum in the charged channel, since the ERT form of the $X(3872)$ wave function, with only neutral $D$ mesons,  is no longer reliable above this momenta. If the hadronic loop integral is performed with a cutoff, $\Lambda$, such that $\gamma_n \ll \Lambda \ll \sqrt{m_D E_\pi }$, one recovers the results from the factorization formalism. We also show that if the theory contains both charged and neutral $D$ mesons and the cutoff is taken to be large  compared to $\sqrt{m_D E_\pi}$,
 and the couplings of the $X(3872)$ to the charged and neutral channels are equal, then the large contributions from the $O(\sqrt{m_D E_\pi })$ part of the hadronic loop integrals cancel and the remainder is well approximated by the factorization formulae, with $\Lambda \approx \pi\gamma_c/2\approx 200$ MeV. In the final section we give our conclusions.

Our study is closely  related to that in Ref.~\cite{Hanhart:2007wa} which compared the wavefunction at the origin squared prescription  to the hadronic loop approach in hadronic molecule decays to two photons. Their main conclusion, relevant to this paper,  is that when the range of the forces binding the hadronic molecule is  much smaller than the distance scale associated with the annihilation, the hadronic loop approach is appropriate, while the wavefunction at the origin prescription is appropriate in the opposite limit. This is consistent with our analysis, but it is unclear whether the assumptions appropriate to the hadronic loop approach apply in the case of the $X(3872)$. The momentum scale characterizing the annihilation process is $\sqrt{m_D E_\pi} \sim 850$ MeV, corresponding to a length scale of $\approx 0.23$ fm, which is comparable to the size of the charmed mesons themselves. It is not clear {\it a priori} that the ERT  wavefunction of  $X(3872)$  will be correct down to such a short distance, but if it is then charged charmed mesons must be included as explicit degrees of freedom. If the ERT wavefunctions are only valid for much longer distance scales, than a factorization approach may be more appropriate. 
Hopefully, future experimental and theoretical studies will clarify which approach is more suitable for $X(3872)$.

\section{Hadronic Loops}

In this section we will consider the $X(3872)$ decays to $\chi_{cJ}$ in the hadronic loops approach. The LO HH$\chi$PT lagrangian for the charmed mesons is 
\bea
{\cal L} &=& {\rm Tr}[H^\dagger_a (i D_0)_{ba} H_b] - g\, {\rm Tr}[H^\dagger_a  H_b \,  \vec{\sigma} \cdot  \vec{A}_{ba}]
+\frac{\Delta_H}{4}{\rm Tr}[H^\dagger_a \,  \sigma^i \, H_a \, \sigma^i] \,  \nn \\
&+&{\rm Tr}[\Hb^\dagger_a (i D_0)_{ab} \Hb_b] + g \,{\rm Tr}[\Hb^\dagger_a   \, \vec{\sigma}\cdot \vec{A}_{ab} \Hb_b]
+\frac{\Delta_H}{4}{\rm Tr}[\Hb^\dagger_a \,  \sigma^i \, \Hb_a \, \sigma^i] \, .
\eea
We use the two component notation of Ref.~\cite{Hu:2005gf}. The field $H_a$ is given by
\bea
H_a =\vec{V}_a \cdot {\vec{\sigma}}+ P_a \, ,
\eea
where $\vec{V}_a$ annihilates $D^*_a$ mesons and $P_a$ annihilates $D_a$ mesons. The subscript
$a$ is an $SU(2)$ index, and $a=1$ for neutral $D$ mesons. The corresponding field for antimesons is $\Hb_a$.
The field ${\vec A}_{ab}$ is the axial current of chiral perturbation theory, $\vec{A}_{ab} = - \vec{\nabla} \pi_{ab}/f_\pi + ...$, 
where $f_\pi$ is the pion decay constant and $\pi_{ab}$ are the Goldstone boson fields.  The lagrangian coupling the $\chi_{cJ}$ to heavy mesons is
\bea\label{lagchi}
{\cal L}_\chi = i\frac{g_1}{2} {\rm Tr}[\chi^{\dagger \, i}  H_a \sigma^i \, \Hb_a] 
+ \frac{c_1}{2} {\rm Tr}[ \chi^{\dagger\, i} H_a \sigma^j \,\Hb_b] \epsilon_{ijk} A^k_{ab}+{\rm h.c.} \, ,
\eea
where the $\chi_{cJ}$ fields 
are represented by 
\bea
\chi^i &=& \sigma^j \, \chi^{ij} \nn \\
&=& \sigma^j\left(\chi^{ij}_2 + \frac{1}{\sqrt{2}}\,\epsilon^{ijk} \chi^k_1 + \frac{\delta^{ij}}{\sqrt{3}} \chi_0 \right)
\, .
\eea
The transformation rules for the various fields under the symmetries of the theory  can be found in  Ref.~\cite{Fleming:2008yn}. 

\begin{figure}[!t]
\begin{center}
\includegraphics[width=15cm]{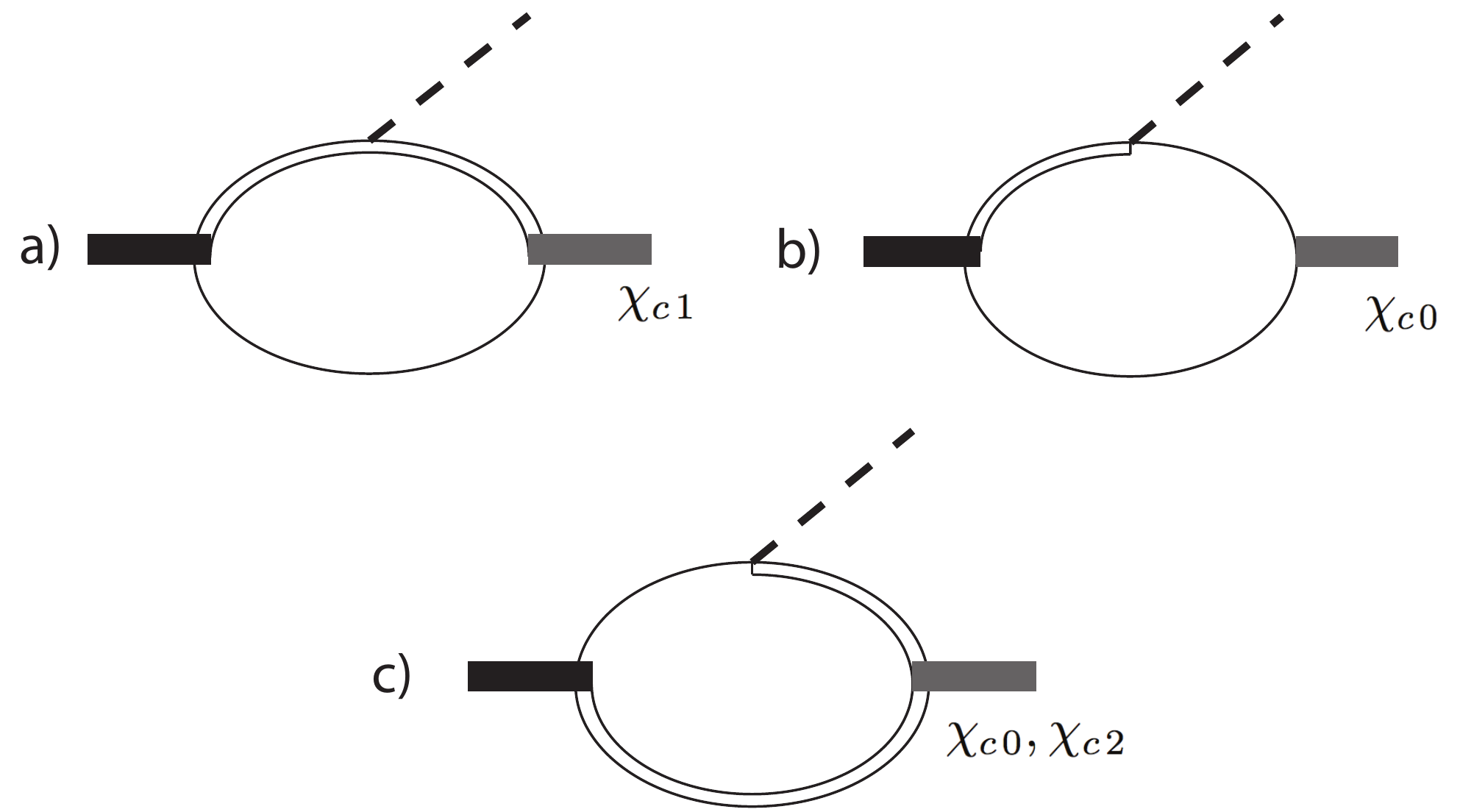}
\end{center}
\vspace{- 0.5 cm}
\caption{
\label{XtoChi} 
\baselineskip 3.0ex
Hadronic loop diagrams contributing to the decays  $X(3872) \to \chi_{cJ}\pi^0$.}
\end{figure}

In the first part of this section we will include only the neutral $D$ mesons  as physical 
degrees of freedom. The hadronic loop diagrams for the decays of $X(3872)$ to the $\chi_{cJ}$ are shown in Fig.~\ref{XtoChi}.
In this figure the black line represents the interpolating field for the  $X(3872)$ and the gray lines are the $\chi_{cJ}$.
The internal lines are the  neutral $D$ mesons, with a single line representing the $D^0$ or $\bar{D}^0$  and a double line for the $D^{*0}$ or $\bar{D}^{*0}$.
 For power counting we use the $v$ counting of Refs.~\cite{Guo:2013zbw,Guo:2014taa}, which is appropriate for the hadronic loop approach. 
The couplings of the $X(3872)$ and the $\chi_{cJ}$ to the $D$ mesons have no derivatives, so these scale as $v^0$. 
The pion is derivatively coupled so that interaction scales as $p_\pi$. In the loops, the integration measure scales as $v^5$ and each propagator 
scales as $v^{-2}$, so the diagrams scale as $p_\pi/v$. There is also a loop diagram that contains a bubble with the four-particle interaction 
 multiplied by $c_1$ in Eq.~(\ref{lagchi}). This diagram contains one fewer propagator, the four-particle  interaction still has a derivative acting on the pion field, so 
the diagram  scales as $p_\pi v$ and is suppressed by $v^2$ in the $v$ expansion. Finally, there is a possible tree-level $X(3872)$-$\chi_{cJ}$-$\pi$ coupling, which would scale as $p_\pi$ and is suppressed by $v$ in the $v$ expansion.
Hence, the diagrams shown in Fig.~\ref{XtoChi} are the leading contribution 
to $X(3872) \to \chi_{cJ} \pi^0$ in the hadronic loop approach.  

 The coupling  of the $X(3872)$ to the $D^0\bar{D}^{*0} + c.c.$ is 
$\sqrt{2\pi \gamma_n/\mu_{DD^*}}$\cite{Weinberg:1962hj}, where $\mu_{DD^*}$ is the reduced mass of the $D^0$ and $\bar{D}^{*0}$ and  $\gamma_n$ is the binding momentum in the neutral channel, i.e.,
$\gamma_n= \sqrt{2\mu_{DD^*} BE_n}$, where $BE_n$ is the binding energy in the neutral channel, $m_{D^{*0}}+m_{D^{0}}-m_{X(3872)}$. If one uses the interpolating field $(D^0 \bar{D}^{*0} +\bar{D}^0 D^{*0}) /\sqrt{2}$ to represent the $X(3872)$ 
this factor arises from wave function renormalization obtained 
using the LSZ formalism for composite operators, see, e.g., Refs.~\cite{Kaplan:1998sz,Fleming:2007rp}. 
Computation of the rates is straightforward and we simply quote the prediction for the rates:
\bea\label{rates}
\Gamma[X(3872) \to \chi_{c0}\pi^0] &=& \frac{ 2g^2 g_1^2}{9 \pi^2 f_\pi^2} \gamma_n\mu_{DD^*}^2 \frac{m_{\chi_{c0}}}{m_{X(3872)}} \,p_\pi^3 \, F_0[\gamma_n,\Delta_0, E_\pi] ^2\nn \\
\Gamma[X(3872) \to \chi_{c1}\pi^0] &=& \frac{g^2 g_1^2}{6 \pi^2 f_\pi^2} \gamma_n\mu_{DD^*}^2 \frac{m_{\chi_{c1}}}{m_{X(3872)}} \, p_\pi^3 \, F_1[\gamma_n,\Delta_0, E_\pi] ^2\nn \\
\Gamma[X(3872) \to \chi_{c2}\pi^0] &=& \frac{5 g^2 g_1^2}{18 \pi^2 f_\pi^2} \gamma_n\mu_{DD^*}^2 \frac{m_{\chi_{c2}}}{m_{X(3872)}} \, p_\pi^3 \, F_2[\gamma_n,\Delta_0, E_\pi]^2 \, .
\eea
Here $g=0.54$ is the axial coupling~\footnote{This value for $g$ is obtained using the recent measurement of $\Gamma[D^{*+}] =83.4 \pm 1.8 $ keV\cite{Lees:2013zna,Agashe:2014kda}  times the measured strong decay  
branching fractions for the $D^{*+}$~\cite{Agashe:2014kda} and the tree-level HH$\chi$PT expression for the strong decay width of the $D^{*+}$. } of the $D$ mesons to the pion, $g_1$ is the coupling of the $\chi_{cJ}$ to $D$ mesons,  $f _\pi=130 $ MeV is the pion decay constant, $m_{\chi_{cJ}} (m_{X(3872)})$ is the mass of the $\chi_{cJ}$ ($X(3872)$),   and $E_\pi(p_\pi)$ is the energy (momentum) of the pion 
in the decay. 
The factors $F_i[\gamma_n,\Delta_0, E_\pi]$ come from the loop integration and are given by
\bea\label{Fs}
F_0[\gamma_n,\Delta_0, E_\pi]  &=& \frac{3}{4}F\left(\gamma_n^2, 2 \mu_{DD^*} (E_\pi-\Delta_0) + \frac{p_\pi^2}{2}, \frac{p_\pi^2}{4}\right) \\
&&+ \frac{1}{4} F\left(\gamma_n^2, 2 \mu_{DD^*} (E_\pi+ \Delta_0)+ \frac{p_\pi^2}{2},\frac{p_\pi^2}{4}\right) \nn \\
F_1[\gamma_n,\Delta_0, E_\pi] &=& F\left(\gamma_n^2, 2 \mu_{DD^*} E_\pi + \frac{p_\pi^2}{2}, \frac{p_\pi^2}{4}\right) \nn \\
F_2[\gamma_n,\Delta_0, E_\pi] &=& F\left(\gamma_n^2, 2 \mu_{DD^*}( E_\pi+\Delta_0) + \frac{p_\pi^2}{2}, \frac{p_\pi^2}{4}\right) \, . \nn
\eea        
To simplify Eq.~(\ref{Fs})  in  some places we have approximated $m_{D^0} \approx m_{D^{*0}} \approx 2 \mu_{DD^*}$, which is accurate to 4\%. In Eq.~(\ref{Fs}),  $\Delta_0 = m_{D^{*0}} - m_{D^0}$ is the hyperfine splitting for the neutral $D$ mesons,  and the function $F(a,b,c)$ is given by
\bea\label{int}
F(a,b,c) &=& \int_0^1 dx \frac{1}{\sqrt{a+b\, x - c\, x^2}} \, .\nn \\
&=& \frac{1}{\sqrt{c}} \left[\tan^{-1}\left(\frac{b}{2\sqrt{a c}}\right) - \tan^{-1} \left( \frac{b-2 c}{2\sqrt{c}\sqrt{a+b-c}} \right)\right] \nn \\
&=&  \frac{1}{\sqrt{c}} \left[\sin^{-1}\left(\frac{b}{\sqrt{b^2+4 a c}}\right) - \sin^{-1} \left( \frac{b-2 c}{\sqrt{b^2+4 ac}} \right)\right] \, .
\eea
The first analytic expression for the evaluation of the integral is appropriate for $a,b,c>0$, $a,c, a+ b-c \neq 0$, which is always the case for us. In our case we always have $b\gg a,c$, and the  
second analytic expression in Eq.~(\ref{int}) is better suited for expanding in $a/b$ and/or $c/b$. In the heavy quark limit  where the $\chi_{cJ}$ are degenerate and $\Delta_0 =0$, $F_0[\gamma_n,\Delta_0, E_\pi] =F_1[\gamma_n,\Delta_0, E_\pi] =F_2[\gamma_n,\Delta_0, E_\pi] $, $E_\pi$ and $p_\pi$ are the same for all three decays,  and the rates are in the ratio
$\Gamma_0: \Gamma_1:\Gamma_2 :: 4:3:5$, where $\Gamma_J  \equiv \Gamma[X(3872) \to \chi_{cJ}\pi^0]$.  In reality, the small hyperfine splittings significantly affect  the value 
of $p_\pi$ multiplying each decay, and the factors of $p_\pi^3 (m_{\chi_{cJ}}/m_{X(3872)}) F_J[\gamma_n,\Delta_0, E_\pi] $ differ significantly between the three decays, so we find 
\bea
\Gamma_0: \Gamma_1:\Gamma_2 :: 3.2: 1.2 : 1.0 \, .
\eea 
Ref.~\cite{Dubynskiy:2007tj} calculates these ratios by weighting the heavy quark spin symmetry prediction with the $p_\pi^3$ factors multiplying each decay, obtaining
$\Gamma_0: \Gamma_1:\Gamma_2 :: 4 p_\pi^3:3 p_\pi^3:5p_\pi^3:: 2.7: 0.95 : 1.0$. The factors of $p_\pi^3$ account for most of the deviation from heavy quark spin symmetry predictions, 
remaining factors give  corrections of order $20-25\%$.

To compute the absolute rates in this approach, one needs to know the coupling constant $g_1$ in Eq.~(\ref{rates}) and the binding momentum, $\gamma_n$.  From the binding energy 
$BE_n =  0.11 \pm 21$ MeV, we find $\gamma_n=14.6^{+12.3}_{-14.6}$ MeV. Because $\chi_{cJ}$ is a conventional quarkonium rather than a bound state of charmed mesons the
coupling $g_1$ is an unknown parameter. We will use the results of Ref.~\cite{Colangelo:2003sa}, which estimates the coupling~\footnote{Our definition of the coupling $g_1$ is a factor of $1/\sqrt{2}$ smaller 
than the $g_1$ defined in Ref.~\cite{Colangelo:2003sa}.}  by  using a vector meson dominance argument to find $g_1^2 \approx m_{\chi_{c0}}/(6 f^2_{\chi_{c0}})$, where $f_{\chi_{c0}} = \langle 0 | \bar{c}c |\chi_{c0}\rangle$ and is calculated to be 510 MeV from QCD sum rules. Using this estimate, $g_1^2 =1/(457 \,{\rm MeV})$, and we find
\bea\label{numrates}
\Gamma[X(3872) \to \chi_{c0}\pi^0] &=&3.8 \, {\rm MeV} \nn \\
\Gamma[X(3872) \to \chi_{c1}\pi^0] &=& 1.4 \, {\rm MeV} \nn \\
\Gamma[X(3872) \to \chi_{c2}\pi^0] &=&  1.2 \, {\rm MeV} \, .
\eea
All of these partial widths separately  exceed the current experimental bound on the {\it total} width, $\Gamma_X < 1.2$ MeV~\cite{Agashe:2014kda}.

The partial widths, $\Gamma[X(3872) \to \chi_{cJ}\pi^0]$, which are presently unmeasured, must in fact be  orders of magnitude smaller than the existing bound on the total width. 
We will next find an upper bound on the sum of the partial widths, 
$\sum_J \Gamma[X(3872) \to \chi_{cJ}\pi^0]$. Theoretical 
calculations~\cite{Voloshin:2003nt,Fleming:2007rp,Baru:2011rs,Guo:2014hqa} of $\Gamma[X(3872) \to D^0 \bar{D}^0 \pi^0]$ find $\Gamma[X(3872) \to D^0 \bar{D}^0 \pi^0] = \Gamma[D^{*0}\to D^0 \pi^0]$ in the limit of zero binding energy. $\Gamma[D^{*0} \to D^0 \pi^0]$ has not been directly measured, but  can be obtained using the total width $\Gamma[D^{*+}] = 83.4 \pm 1.8$ keV~\cite{Agashe:2014kda,Lees:2013zna} and ${\rm Br}[D^{*+} \to D^+ \pi^0] = 30.7 \pm 0.5 \%$ which together give $\Gamma[D^{*+} \to D^+ \pi^0] = 25.6 \pm 0.69$ keV. In the isospin symmetry limit, $\Gamma[D^{*+} \to D^+ \pi^0] = \Gamma[D^{*0}\to D^0 \pi^0]$. Noting that each decay scales like $p_\pi^3$ and taking into account differences in phase space, we find $\Gamma[D^{*0}\to D^0 \pi^0] = 36.4 \pm 0.98$ keV. Therefore, we expect  $\Gamma[X(3872) \to D^0 \bar{D}^0 \pi^0]   = 36^{+6}_{-10}$ keV.  The central value here is our extracted value of $\Gamma[D^{*0}\to D^0 \pi^0]$, which has only a few percent uncertainty
from experimental uncertainties and isospin violation. The uncertainty  in $\Gamma[X(3872) \to D^0 \bar{D}^0 \pi^0]$ is obtained by assuming that the binding energy of the $X(3872)$ is between 0 and 0.3 MeV, and using the theoretical calculation of $\Gamma[X(3872) \to D^0 \bar{D}^0 \pi^0]$ in Ref.~\cite{Fleming:2007rp}, which includes corrections from range corrections, pion loops, and higher dimension operators.  Furthermore, the branching fraction $\Gamma[X(3872) \to D^0 \bar{D}^0 \pi^0]/\Gamma[X(3872)] > 32\%$~\cite{Agashe:2014kda}, implying  $\Gamma[X(3872)] \lesssim 131$ keV. (We use the largest value of $\Gamma[X(3872) \to D^0 \bar{D}^0 \pi^0]$ in our quoted range to obtain this bound.) 
The branching ratio for any of the strong decays to conventional quarkonia is considerably smaller than this. For example, 
$\Gamma[X(3872)\to D^0\bar{D}^{0}\pi^0]/\Gamma[X(3872) \to J/\psi \pi^+ \pi^-] = 8.8 ^{+3.1}_{-3.6}$~\cite{Agashe:2014kda}, implying $\Gamma[X(3872) \to J/\psi \pi^+ \pi^-] = 4.1^{+2.8}_{-1.1}$ keV if $\Gamma[X(3872) \to D^0 \bar{D}^0 \pi^0] = 36$ keV. The total partial width to final states $D^0\bar{D}^0\pi^0, J/\psi \pi^+ \pi^-, J/\psi \pi^+\pi^-\pi^0$, and $\psi(2S) \gamma$ constitute at least 39.5\% of the total 
width~\cite{Agashe:2014kda}, so the total partial width to all other states is less than  79 keV, so  $\sum_J \Gamma[X(3872) \to \chi_{cJ}\pi^0] < 79$ keV and we can see from Eq.(\ref{numrates}) that the hadronic loops prediction
is almost two orders of magnitude too large. If this situation is to be fixed by using smaller values of $g_1^2$ and changing no other parameters,  we must require $g_1^2 \leq 1/(37\, {\rm GeV})$ which seems implausibly small from the point of view of naive dimensional analysis. 

One way to fix this is to include both the charged and neutral mesons in the theory, since the decay rate is naturally suppressed if the $X(3872)$ couples to charm-anticharm mesons in the $I=0$ channel. Refs.~\cite{Aceti:2012cb,Aceti:2012qd} have emphasized the necessity of including both charged and neutral $D$ mesons in the context of $X(3872) \to J/\psi\gamma, J\psi \pi^+ \pi^- $ and $J/\psi \pi^+ \pi^- \pi^0$ decays. When the charged channel is included as well the formulae of Eq.~(\ref{rates}) generalize to  
\bea 
\Gamma[X(3872) \to \chi_{c0}\pi^0] &=& \frac{ g^2 g_1^2}{9\pi^3 f_\pi^2}  \mu_{DD^*}^4\frac{m_{\chi_{c0}}}{m_{X(3872)}} \,p_\pi^3 \, 
\left( g_0  F_0[\gamma_n,\Delta_0, E_\pi] - g_+ F_0[\gamma_c,\Delta_+, E_\pi] \right)^2   \\
\Gamma[X(3872) \to \chi_{c1}\pi^0] &=& \frac{g^2 g_1^2}{12 \pi^3 f_\pi^2} \mu_{DD^*}^4\frac{m_{\chi_{c1}}}{m_{X(3872)}} \, p_\pi^3 \, 
( g_0 F_1[\gamma_n,\Delta_0, E_\pi] -g_+  F_1[\gamma_c,\Delta_+, E_\pi] )^2\nn \\
\Gamma[X(3872) \to \chi_{c2}\pi^0] &=& \frac{5 g^2 g_1^2}{36\pi^3 f_\pi^2} \mu_{DD^*}^4 \frac{m_{\chi_{c2}}}{m_{X(3872)}} \, p_\pi^3 \,
(g_0  F_2[\gamma_n,\Delta_0, E_\pi] - g_+  F_2[\gamma_c,\Delta_+, E_\pi] ) ^2 \, , \nn
\eea
where $\gamma_c$ is the binding momentum in the charged channel, 
$\gamma_c= \sqrt{2\mu_{DD^*} BE_c}$, where $BE_c = m_{D^{*\pm}}+m_{D^{\pm}} -m_{X(3872)}$, and 
 $g_0$ and $g_+$ are the  couplings of the $X(3872)$ to the neutral  and charged channels. These obey the constraint \cite{Gamermann:2009uq}
\bea\label{wcc}
g_0^2 \, {\rm Re}\Sigma_0'(-E_X) + g_+^2 \, {\rm Re}\Sigma_+'(-E_X) =1.  
\eea
where  $\Sigma_0(-E_X)$ and $\Sigma_+(-E_X)$ are the contribution to the self-energy of the $X(3872)$  from the neutral and charged mesons,
respectively, and $'$ denotes differentiation with respect to the energy. Eq.~(\ref{wcc}) can derived by solving the coupled channel problem, see for example
Ref.~\cite{Mehen:2011yh} where the coupled channel problem is solved for a theory of non-relativistic heavy mesons with contact interactions 
that mediate $S$-wave scattering in both the $I=0$ and $I=1$ channels. The coupling 
can be extracted from the residues of the $T$-matrix at the $X(3872)$ pole, which can be shown to satisfy Eq.~(\ref{wcc}).\footnote{I thank
R.~P.~Springer and J.~Z.~Liu for discussions about this point.} If only $I=0$ scattering is present then $g_0=g_+$. 

Noting that ${\rm Re} \, \Sigma_0'(-E_X)= \frac{\mu_{DD^*}^2}{2\pi\gamma_n}$ and ${\rm Re} \,\Sigma_+'(-E_X)= \frac{\mu_{DD^*}^2}{2\pi\gamma_c}$, the 
constraint in Eq.~(\ref{wcc}) can be solved by setting
\bea
g_0 = \sqrt{ \frac{2\pi\gamma_n}{\mu_{DD^*}^2}} \cos \theta \, , \, g_+ = \sqrt{ \frac{2\pi\gamma_c}{\mu_{DD^*}^2}} \sin \theta \, ,
\eea
so the decay rates in terms of $\theta$ and the binding momenta are
\bea 
\Gamma[X(3872) \to \chi_{c0}\pi^0] &=& \\ 
&&\hspace{-0.75 in}\frac{ 2g^2 g_1^2}{9 \pi^2 f_\pi^2}  \mu_{DD^*}^2\frac{m_{\chi_{c0}}}{m_{X(3872)}} \,p_\pi^3 \,
\left(  \cos \theta \sqrt{\gamma_n} F_0[\gamma_n,\Delta_0, E_\pi] -  \sin \theta \sqrt{\gamma_c} F_0[\gamma_c,\Delta_+, E_\pi] \right)^2  \nn \\
\Gamma[X(3872) \to \chi_{c1}\pi^0] &=& \nn \\
&&\hspace{-0.75 in}\frac{g^2 g_1^2}{6 \pi^2 f_\pi^2} \mu_{DD^*}^2 \frac{m_{\chi_{c1}}}{m_{X(3872)}} \, p_\pi^3 \, 
( \cos \theta \sqrt{\gamma_n}  F_1[\gamma_n,\Delta_0, E_\pi] -  \sin \theta \sqrt{\gamma_c}  F_1[\gamma_c,\Delta_+, E_\pi] )^2\nn \\
\Gamma[X(3872) \to \chi_{c2}\pi^0]  &=& \nn  \\
&&\hspace{-0.75 in} \frac{5 g^2 g_1^2}{18 \pi^2 f_\pi^2} \mu_{DD^*}^2 \frac{m_{\chi_{c2}}}{m_{X(3872)}} \, p_\pi^3 \,
(\cos \theta \sqrt{\gamma_n}   F_2[\gamma_n,\Delta_0, E_\pi] -  \sin \theta \sqrt{\gamma_c} F_2[\gamma_c,\Delta_+, E_\pi] ) ^2 \, . \nn
\eea
The actual value of $\theta$ depends on the underlying dynamics, and cannot be determined from the EFT {\it a priori}, so we will leave it as a free parameter. 
By tuning $\theta$ we can arrange a cancellation between charged and neutral loops which allows the prediction to be consistent with the 
bounds. Demanding $\sum_J \Gamma[X(3872) \to \chi_{c J} \pi^0] < 79$ keV, we find that $\theta =0.37\pm 0.04$. 
For this range of $\theta$, $0.78<  g_0/g_+  < 0.99$, so the ratio of these  couplings is close to 1. This range is consistent with scattering being  
dominated by the $I=0$ channel. The constraint on $\theta$, and hence $g_0/g_+$, is correct so long as $\gamma_n\approx 14$ MeV, and 
$g_1\approx 1/(500 \, {\rm MeV})$. Unfortunately, the uncertainties on both these parameters are $O(1)$. If these parameters are an order of magnitude
smaller, which seems unlikely but is not ruled out by experiment, then the constraints on $\theta$ and  $g_0/g_+$ would be considerably weaker.

\begin{figure}[!t]
\begin{center}
\includegraphics[width=15cm]{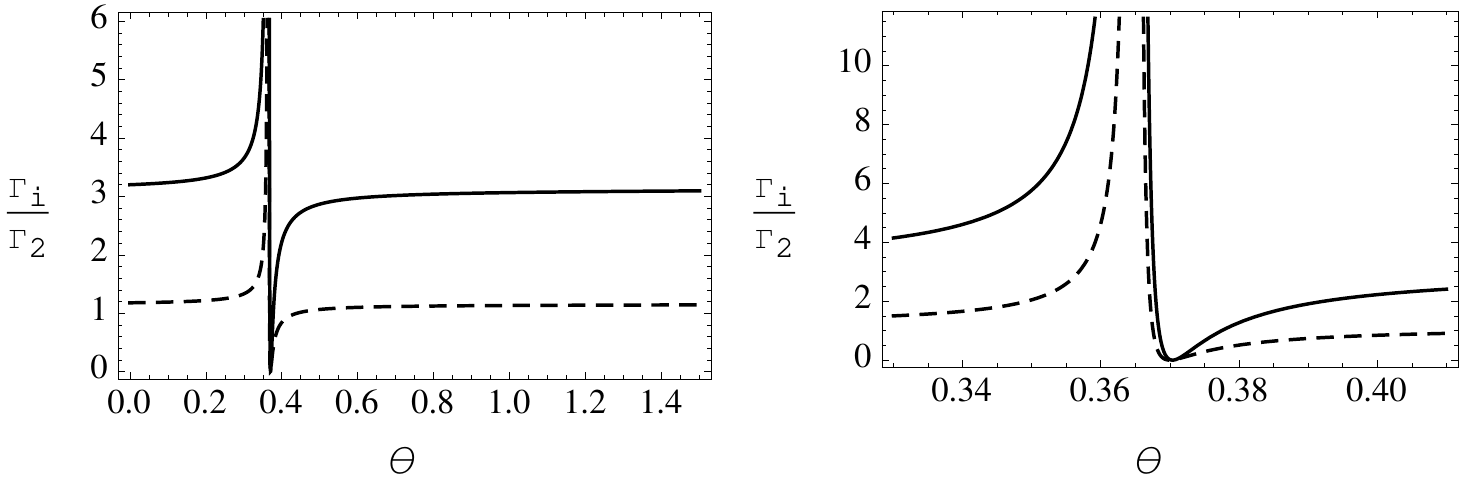}
\end{center}
\vspace{- 0.5 cm}
\caption{
\label{ratio} 
\baselineskip 3.0ex
$\Gamma_0/\Gamma_2$ (solid) and $\Gamma_1/\Gamma_2$ (dashed) as a function of $\theta$.}
\end{figure}

Finally, we comment on the predicted ratios for $\Gamma_0:\Gamma_1:\Gamma_2$ in this approach. Because the desired rates are achieved by a
fine-tuned cancellation between charged and neutral pion loops, the ratios vary wildly as a function of $\theta$ near $\theta = 0.37$ where all three decay rates are  very close to 
zero. The plots in Fig.~\ref{ratio} show the ratios $\Gamma_0/\Gamma_2$ (solid) and $\Gamma_1/\Gamma_2$ (dashed) as a function of $\theta$. The plot on the left in Fig.~\ref{ratio} shows
these ratios for a wide range of $\theta$ and one sees that  $\Gamma_0/\Gamma_2\approx 3.2$ and $\Gamma_1/\Gamma_0\approx 1.2$ for most values of 
$\theta$, except near $\theta=0.37$. The plot on the right shows the prediction for the allowed range $0.33 < \theta < 0.41$.
In this range the ratios deviate significantly from $\Gamma_0:\Gamma_1:\Gamma_2$::3.2:1.2:1. It would be interesting to obtain experimental information on $\Gamma_0:\Gamma_1:\Gamma_2$ as this could distinguish between the various approaches to calculating the $X(3872)$ decays to conventional charmonia. In the hadronic loop approach, with both charged and neutral mesons included as explicit degrees of freedom, measurement of these ratios could determine the correct value of $\theta$.

\section{Factorization}

In this section we discuss how the hadronic loop approach discussed in the previous section is related to the 
factorization approach of Ref.~\cite{Fleming:2008yn}.  We begin by considering the amplitude from the loop diagram in Fig.~\ref{XtoChi}b),~\footnote{The discussion that follows applies to all three diagrams.} with only neutral $D$ mesons in the loop. This is given by
\bea
{\cal A}_{1b}[X(3872) \to \chi_{c0} \pi^0] &=&
i \epsilon_X \cdot p_\pi \sqrt{\frac{3}{2}} \frac{g g_1}{f_\pi} \times \\
&& \hspace{-1.0 in} \int \frac{d^4l}{(2\pi)^4} \frac{1}{E_X - \Delta_0+l_0  -\frac{l^2}{2m_D^*}+i\epsilon} \frac{1}{-l_0- \frac{l^2}{2 m_D}+i\epsilon} \frac{1}{E_X+l_0-E_\pi -\frac{(l-p_\pi)^2}{2m_D} +i\epsilon} \, . \nn
\eea
Here $E_X$ is the energy of the $X(3872)$ relative to $2 m_D$, so $E_X = \Delta_0-\gamma_n^2/(2\mu_{DD^*})$.
The $l_0$ integral is done by contour integration, resulting in the integral:
\bea\label{int1}
{\cal A}_{1b}[X(3872) \to \chi_{c0} \pi^0]  = - \epsilon_X \cdot p_\pi \sqrt{\frac{3}{2}} \frac{g g_1}{f_\pi} \int \frac{d^3l}{(2\pi)^3} \frac{2\mu_{DD^*}}{l^2+\gamma_n^2}\frac{1}{E_X - E_\pi -\frac{l^2+(l-p_\pi)^2}{2m_D}} \, .
\eea
Note that the integrand scales as $\int d^3l/(l^2)^2$ for large $l$ and hence the integral is finite.
When we multiply this amplitude by the factor $\sqrt{2 \pi \gamma_n}/\mu_{DD^*}$ coming from the wavefunction renormalization, this result can be written as 
\bea 
{\cal M}_{1b}[X(3872) \to \chi_{c0} \pi^0]  &=&\frac{\sqrt{2\pi \gamma_c}}{\mu_{DD^*}} {\cal A}_{1b}[X(3872) \to \chi_{c0} \pi^0]   \\
&=&\int\frac{d^3l}{(2\pi)^3} \psi_{DD^*}(\vec{l}\,\,) \,{\cal M}[D^{*0}(\vec{l}\,\,) \bar{D}^{0}(-\vec{l}\,\,) \to \chi_{c0}\pi^0] \, , \nn
\eea 
where 
\bea
\psi_{DD^*}(\vec{l}\,\,) = \frac{\sqrt{8 \pi \gamma_n}}{l^2+\gamma_n^2} \, , 
\eea
is the momentum space wavefunction of the $D^0$-$\bar{D}^{*0}$ in the $X(3872)$, and 
\bea
{\cal M}[D^{*0}(\vec{l}\,\,) \bar{D}^{0}(-\vec{l}\,\,) \to \chi_{c0}\pi^0] &=&- \epsilon_X \cdot p_\pi  \sqrt{\frac{3}{2}} \frac{g g_1}{f_\pi}\frac{1}{E_X -E_\pi - \frac{l^2}{2 m_D} -  \frac{(l-p_\pi)^2}{2 m_D} } \, , 
\eea
is  a  tree-level contribution to the  HH$\chi$PT amplitude for $D^0\bar{D}^{*0} \to \chi_{c0}\pi^0$. The momentum space wavefunction, $\psi_{DD^*}(\vec{l}\,)$, has the form dictated by  ERT and is correct so long as the $D$ mesons  are separated by large distances compared to the strong force that binds them. In a theory in which the 
charged $D$ mesons have been integrated out, the scale $\gamma_c = 126$ MeV should be considered large and the wavefunction $\psi_{DD^*}(\vec{l}\, \,)$ can only be trusted 
below this momentum. 

With this in mind, we will continue evaluating Eq.~(\ref{int1}) in the hadronic loops formalism, but now imposing a UV cutoff on the integral. Combining the two terms with 
Feynman parameters, we get
\bea\label{cutoff}
{\cal A}_{1b}[X(3872) \to \chi_{c0} \pi^0]  &=&\epsilon_X \cdot p_\pi \sqrt{\frac{3}{2}} \frac{g g_1}{f_\pi} \int_0^1 dx   \int^\Lambda \!\! \! \frac{d^3l}{(2\pi)^3} \frac{2\mu_{DD^*} m_D}{ (l^2 + \Delta(x))^2} \\
&&\hspace{-1.0 in}=  \epsilon_X \cdot p_\pi \sqrt{\frac{3}{2}} \frac{g g_1}{4\pi^2f_\pi} 2\mu_{DD^*} m_D\int_0^1 dx \left[-\frac{\Lambda}{\Lambda^2+ \Delta(x)} + \frac{1}{\sqrt{\Delta(x)}} \tan^{-1}\left(\frac{\Lambda}{\sqrt{\Delta(x)}} \right) \right]\, , \nn
\eea 
where $\Delta(x)$ is given by
\bea
\Delta(x) = \gamma_n^2 + x\left(m_D(E_\pi -\Delta_0) +\frac{p_\pi^2}{2} +\gamma_n^2 \frac{m_D-2\mu_{DD^*}}{2 \mu_{DD^*}}\right )- x^2 \frac{p_\pi^2}{4} \, . 
\eea 
The term $\gamma_n^2 (m_D-2\mu_{DD^*})/(2 \mu_{DD^*}) \approx -2.8\,{\rm MeV}^2$ is negligible compared to the remaining terms so we will drop it
as well as  the terms proportional to $p_\pi^2$ in $\Delta(x)$ since $p_\pi^2 \ll m_D(E_\pi- \Delta_0)$. One can check that setting $p_\pi^2=0$ only 
changes the numerical values of the functions in Eq.~(\ref{Fs}) by a few percent.  This approximation allows the integral in Eq.~(\ref{cutoff}) to be evaluated 
analytically and one obtains 
\bea
{\cal A}_{1b}[X(3872) \to \chi_{c0} \pi^0]  &=&\epsilon_X \cdot p_\pi \sqrt{\frac{3}{2}} \frac{g g_1}{2 \pi^2 f_\pi} \frac{2\mu_{DD^*}}{E_\pi -\Delta_0} \\
&&\hspace{-1.0 in} \times \left[  \sqrt{m_D(E_\pi-\Delta_0)+ \gamma_n^2} \tan^{-1}\left( \frac{\Lambda}{\sqrt{ m_D(E_\pi-\Delta_0)+\gamma_n^2}}\right)- \gamma_n \tan^{-1}\left( \frac{\Lambda}{\gamma_n}\right)\right] \nn \,.
\eea
Since the integral is finite we can send $\Lambda\to \infty$ and the result is 
\bea\label{LamInf}
{\cal A}_{1b}[X(3872) \to \chi_{c0} \pi^0] = \epsilon_X \cdot p_\pi \sqrt{\frac{3}{2}} \frac{g g_1}{4 \pi f_\pi} \frac{2\mu_{DD^*}}{E_\pi -\Delta_0} 
\left[ - \gamma_n+ \sqrt{m_D(E_\pi-\Delta_0)+ \gamma_n^2}\right]   \,.
\eea
This is the result from the hadronic loops formalism when we set $p_\pi^2=0$. To see this it is helpful to note $F(a,b,0) = 2(- \sqrt{a}+\sqrt{a+b})/b$. As stated earlier, setting $p_\pi^2=0$ is an excellent approximation to the exact result. 
 However, we have argued that in a theory without explicit charged mesons the 
cutoff $\Lambda$ should be not much larger than $\gamma_c \approx 126$ MeV. The factor  $m_D(E_\pi -\Delta_0) \approx (736\, {\rm MeV})^2$, so 
$m_D(E_\pi -\Delta_0)  \gg  \gamma_c^2$. In Figs.~\ref{XtoChi}a) and c), the factor of $m_D(E_\pi -\Delta_0)$ is replaced with $m_{D^{(*)}} (E_\pi+\Delta_0)$
or  $2 \mu_{DD^*} E_\pi$, which are even larger.  Typically these quantities are of order $(850\,{\rm MeV})^2$ and can be as large as $(1073 \,{\rm MeV})^2$. 
So, for a physical value of the cutoff  we should take $\gamma_n \ll \Lambda \ll \sqrt{m_D (E_\pi-\Delta)}$, then we have 
\bea\label{lowLam}
{\cal A}_{1b}[X(3872) \to \chi_{c0} \pi^0]&=& \epsilon_X \cdot p_\pi \sqrt{\frac{3}{2}} \frac{g g_1}{4\pi f_\pi} \frac{2\mu_{DD^*}}{E_\pi -\Delta_0} \left(\frac{2\Lambda}{\pi}-\gamma_n \right)
+  O\left( \frac{\gamma_n}{\Lambda}, \frac{\Lambda}{m_D E_\pi}\right).
\eea

This is actually the result in the factorization approach. Starting with Eq.~(\ref{int1}) we note that in the second propagator $E_X - E_\pi -  (l^2+(l-p_\pi)^2)/(2m_D) \approx \Delta_0-E_\pi + O(\frac{Q^2,p_\pi Q, p_\pi^2}{m_D E_\pi})$,
where $l \sim \gamma_n \sim Q$ ($Q$ denotes any generic scale of order the binding momentum).  
Consistency of XEFT power counting requires that we drop the $O(\frac{Q^2,p_\pi Q, p_\pi^2}{m_D E_\pi})$ terms. 
 Then the $l$ integral is straightforward and one obtains Eq.~(\ref{lowLam}) for the amplitude. Note that after dropping the $O(\frac{Q^2,p_\pi Q, p_\pi^2}{m_D E_\pi})$ terms
 the integrand scales as $\int d^3l/l^2$ for large $l$, hence the integral is divergent and depends on the cutoff.
In the factorization formalism, the divergent integral is interpreted as the nonperturbative matrix element
\bea\label{ldme}
 \frac{1}{3}\sum_\lambda |\langle 0| \frac{1}{\sqrt{2}}{\epsilon}_i(\lambda) 
\,(V^i \, \Pb +\Vb^i \, P) |X(3872,\lambda)\rangle|^2  &=& \frac{  \gamma_n}{2\pi } \left(\frac{2\Lambda}{\pi}-\gamma_n\right)^2  . 
\eea 
Here the evaluation of this matrix element is sensitive to the cutoff $\Lambda$. This indicates the matrix element is sensitive to the short-distance nature 
of the $X(3872)$ and cannot be calculated with XEFT. Still, we can use the formula in Eq.~(\ref{ldme}) to parametrize the matrix element and the constraint on this matrix element 
from the requirement $\sum_J \Gamma[X(3872) \to \chi_{cJ} \pi^0] < 79$ keV, when expressed in terms of $\Lambda$, is $\Lambda \leq 325$ MeV. This confirms that in the theory 
with charged mesons integrated out, the cutoff must be interpreted as being a few hundred MeV at most and much lower than the scale set by  $\sqrt{m_D E_\pi }$.

In the context of Non-Relativistic QCD, making similar expansions in non-relativistic propagators inside loop diagrams in order  to maintain consistent power counting is known as the multipole expansion~\cite{Grinstein:1997gv}. In the present case, this keeps contributions from the loop integral that come from low scales $l\sim \gamma_n$ but discards contributions that come from high momentum region of integration $l \sim \sqrt{ m_D E_\pi }\sim 850$ MeV. When the cutoff is taken to infinity, contributions 
from both regions contribute to the finite answer (see the two terms in Eq.~(\ref{LamInf}) ) and the contributions from large $l\sim \sqrt{ m_D E_\pi }$ give the dominant contribution. 
It is conceivable that in a theory with explicit charged and neutral $D$ mesons the true cutoff can be taken $O(GeV)$ and this second contribution can be reliably computed. 
But in a theory with only neutral $D$ mesons the cutoff cannot be interpreted as being much higher $\gamma_c \sim 126$ MeV, otherwise the charged mesons should appear as explicit degrees of freedom. 

Finally we  consider what happens when the charged mesons are included in the theory. Let us assume that in the theory with explicit charged mesons that we can take $\Lambda$ to be large and keep the region of the integral from $l \sim \sqrt{m_D E_\pi }$.
Neglecting terms suppressed by
$p_\pi^2/(m_D E_\pi)$, the contribution to the matrix element for $X(3872) \to \chi_{c0} \pi^0$ from the  diagram in Fig.~\ref{XtoChi}b), with both neutral  and charged mesons and the relevant 
couplings included, is   
\bea
{\cal M}_{1b}[X(3872) \to \chi_{c0} \pi^0] &=& \epsilon_X \cdot p_\pi \sqrt{\frac{3}{2}} \frac{g g_1}{4 \pi f_\pi} 2\mu_{DD^*}  \\
&&\hspace{-1.0 in}\times
\left[g_0 \frac{- \gamma_n + \sqrt{m_D(E_\pi-\Delta_0)+ \gamma_n^2}}{E_\pi-\Delta_0} +g_+ \frac{ \gamma_c -  \sqrt{m_D(E_\pi-\Delta_+)+ \gamma_c^2}}{E_\pi-\Delta_+} \right]  \nn \,.
\eea
 In the $I=0$ limit, $g_0 = g_+$, the terms proportional to $g_0 \sqrt{m_D(E_\pi-\Delta_0)+ \gamma_n^2}$ and $g_+ \sqrt{m_D(E_\pi-\Delta_+)+ \gamma_c^2}$ essentially cancel because they differ by only 2\% in magnitude. Noting that $\Delta_0/\Delta_+ = 1.01$, the final result is well approximated by 
\bea
{\cal M}_{1b}[X(3872) \to \chi_{c0} \pi^0] &=&g_0  \epsilon_X \cdot p_\pi \sqrt{\frac{3}{2}} \frac{g g_1}{4 \pi f_\pi} \frac{2\mu_{DD^*}}{E_\pi -\Delta_0}   \left[- \gamma_n + \gamma_c \right]  \,.
\eea
which is the factorization result in a theory with only neutral mesons, Eq.~(\ref{lowLam}), with the UV cutoff, $\Lambda$, replaced with $\Lambda= \pi\gamma_c/2 \approx  198$ MeV. 
So in this limit  the hadronic loops result is equal to the factorization result in a theory with only neutral $D$ mesons, with an appropriately  low value  for the UV cutoff. Note that in isospin conserving decays the high energy contributions from charged and neutral loops will add rather than cancel. Their effects must be reproduced by diagrams with local counterterms in XEFT.

\section{Conclusions}
 
 In this paper we studied the decays $\Gamma[X(3872) \to \chi_{cJ} \pi^0]$ within the two commonly used approaches to calculating $X(3872)$ decays to conventional quarkonium within EFT: the hadronic loop approach and the factorization approach. Within the hadronic loop approach,  we find that  if one only includes neutral mesons as explicit degrees of freedom, and uses the estimate of the $\chi_{cJ}$ coupling to $D$ mesons from Ref.~\cite{Colangelo:2003sa}, then predictions for each of these partial widths exceeds the known bound on the {\it total width}. We then obtained a bound on $\sum_J\Gamma[X(3872) \to \chi_{cJ} \pi^0]$ by exploiting the fact that $\Gamma[X(3872) \to D^0 \bar{D}^0 \pi^0] = \Gamma[D^{*0}\to D^0\pi^0]$ in the limit of small binding energy within ERT. Combining this with known results for $\Gamma[X(3872) \to D^0 \bar{D}^0 \pi^0]/\Gamma[X(3872)]$ as well as lower bounds on branching fractions to observed decays  of the $X(3872)$, we found $\sum_J\Gamma[X(3872) \to \chi_{cJ} \pi^0] < 79$ keV. To calculate the  theoretical uncertainties in the estimation of $\Gamma[X(3872) \to D^0 \bar{D}^0 \pi^0]$ we use the results of Ref.~\cite{Fleming:2007rp}, which supplement ERT with range corrections, corrections from higher dimension operators in XEFT, and pion exchange. We conclude that the prediction for $\sum_J\Gamma[X(3872) \to \chi_{cJ} \pi^0]$ is almost two orders of magnitude too large if $\gamma_n\approx 14$ MeV and $g_1 \approx 1/(457 \, {\rm MeV})$.  Within the hadronic loop approach, the prediction for $\sum_J\Gamma[X(3872) \to \chi_{cJ} \pi^0]$ can be made consistent with experiment by including charged charmed mesons in addition to neutral charmed mesons as explicit degrees of freedom. The couplings of the $X(3872)$ to the neutral ($g_0$)  and charged ($g_+$) channels must be tuned to arrange a near cancellation between the charged and neutral meson loop contributions. Consistency with data requires $0.78 <g_0/g_+< 0.99$. If $X(3872)$ appeared as a pole in the $I=0$ channel only then we would expect $g_0/g_+ =1$, so the cancellation is naturally explained if the $X(3872)$ is an $I=0$ state.
 
 Next we discussed the relationship between the hadronic loop approach and the factorization approach to $X(3872)$ decays. We showed that the hadronic loop diagram is proportional to the integral $\int d^3l \,\psi_{DD^*}( l) {\cal M}[D^{*0}(l) \bar{D}^0(-l) \to \chi_{cJ} \pi^0]$, where $\psi_{DD^*}(l)$ is the momentum 
 space wave function of the $X(3872)$ predicted by ERT and ${\cal M}[D^{*0}(l) \bar{D}^0(-l) \to \chi_{cJ} \pi^0]$ is the tree-level amplitude for $D^{*0}\bar{D}^0 \to \chi_{cJ}\pi^0$ in HH$\chi$PT. The integrals are well-approximated (within $\sim 5\%$) dropping terms that are $p_\pi^2/(m_D E_\pi)$ suppressed. Making this approximation, we see that the hadronic loop integral contains two widely separated energy scales: $\gamma_n =14$ MeV and $\sqrt{m_D E_\pi}\approx 850$ MeV. The hadronic loop   result is numerically dominated  by large loop momenta of order $\sqrt{m_D E_\pi}$. For these high momenta, $\psi_{DD^*}( l)$ is likely to deviate from ERT form, since this form is only known to be correct for $l \sim \gamma_n$.  If charged mesons have been integrated out of the theory, the ERT form of the wave function is only reliable for $l \lesssim \gamma_c$. If this is the case the theory must be interpreted as having a UV cutoff $\sim 100-200$ MeV. In the limit $\gamma_n \ll \Lambda \ll m_D E_\pi$, the hadronic loop is well approximated by the factorization formulae for the decay rate. 
 
The factorization formulae for the $X(3872)$ decay rate can be interpreted as performing the multipole expansion on the hadronic loop integral, i.e., the XEFT power counting $l \sim \gamma_n \sim Q \ll \sqrt{m_D E_\pi}$ is imposed at the level of the integrand in XEFT.  Since $l \sim \gamma_n \sim Q$ there is no further approximation for $\psi_{DD^*}( l)$, but  within ${\cal M}[D^{*0}(l) \bar{D}^0(-l) \to \chi_{cJ} \pi^0]$ we drop corrections 
 suppressed by $(l,\gamma_n)/m_D$ and $p_\pi/m_D$.   This has the effect of removing the $l \sim \sqrt{m_D E_\pi}$ 
 contribution to the hadronic loop and keeping only the low energy $l \sim \gamma_n$ contributions. In the hadronic loops approach with both charged and neutral charmed meson, 
if we choose $g_0=g_+$  so the large $l \sim \sqrt{m_D E_\pi}$ contributions cancel, the remaining terms in the integral are the same as the factorization result in a theory with explicit neutral mesons only, and $\Lambda = \pi \gamma_c/2 \approx 200$ MeV.
 
 We conclude that within the hadronic loop approach it is inconsistent to keep only neutral charmed mesons and integrate loop momenta to arbitrarily large momentum. If loop integrations are taken to infinity, keeping large contributions from $l \sim m_D E_\pi \sim 850$ MeV then charged $D$ mesons must be included and the coupling of the $X(3872)$ to the charged channel must be nearly equal to that of the neutral channel so the predicted rates for $\sum_J\Gamma[X(3872) \to \chi_{cJ} \pi^0]$ are consistent with data. If the charged mesons are integrated out of the theory, then the cutoff on the loop momenta should be $\Lambda \sim O(\gamma_c)$ and the results of the hadronic loop approach will be consistent with what is obtained in the factorization approach to $X(3872)$ decays to conventional charmonia.
 
\acknowledgments 

We thank H.~Greisshammer, C.~Hanhart, and R.~Springer for discussions pertaining to this work. 
I thank S.~Fleming and J.~Powell for  proofreading this manuscript and I thank C.~Hanhart 
for his many comments on the first version of this paper.
This work was supported in part by the  Director, Office of Science, Office of Nuclear 
Physics, of the U.S. Department of Energy under Contract No.   
DE-FG02-05ER41368.



\end{document}